\documentstyle{article}

\catcode`\@=11
\def\relaxnext@{\let\next\relax}
\font\tenmsy=msym10 scaled\magstep1
\font\sevenmsy=msym7 scaled\magstep1
\font\fivemsy=msym5  scaled\magstep1
\newfam\msyfam
\textfont\msyfam=\tenmsy
\scriptfont\msyfam=\sevenmsy
\scriptscriptfont\msyfam=\fivemsy
\font\teneuf=eufm10 scaled\magstep1
\font\seveneuf=eufm7 scaled\magstep1
\font\fiveeuf=eufm5 scaled\magstep1
\newfam\euffam
\textfont\euffam=\teneuf
\scriptfont\euffam=\seveneuf
\scriptscriptfont\euffam=\fiveeuf
\def\frak{\relaxnext@\ifmmode\let\next\frak@\else
 \def\next{\Err@{Use \string\frak\space only in math mode}}\fi\next}
\def\goth{\relaxnext@\ifmmode\let\next\frak@\else
 \def\next{\Err@{Use \string\goth\space only in math mode}}\fi\next}
\def\frak@#1{{\frak@@{#1}}}
\def\frak@@#1{\noaccents@\fam\euffam#1}
\def\Bbb{\relaxnext@\ifmmode\let\next\Bbb@\else
 \def\next{\Err@{Use \string\Bbb\space only in math mode}}\fi\next}
\def\Bbb@#1{{\Bbb@@{#1}}}
\def\Bbb@@#1{\noaccents@\fam\msyfam#1}
\def\accentfam@{7}
\def\noaccents@{\def\accentfam@{0}}
\catcode`\@=\active

\newcommand{\bz}{{\Bbb Z}}

\newcommand{\bc}{{\Bbb C}}

\makeatletter
\@addtoreset{equation}{section}
\makeatother

\newtheorem{thm}{Theorem}[section]
\newtheorem{prop}[thm]{Proposition}
\newtheorem{lem}[thm]{Lemma}

\begin{document}
\begin{flushright}
RIMS-99? \\
hep-th/9501nnn \\
Jan 1995
\end{flushright}

\vspace{36pt}

\begin{center}
\begin{Large}
Annihilation poles of a Smirnov-type integral formula for solutions \\
to quantum Knizhnik--Zamolodchikov equation

\vspace{24pt}
\end{Large}

\begin{large}
Takeo Kojima\raisebox{2mm}{\scriptsize 1},
        Kei Miki\raisebox{2mm}{\scriptsize 2}
    and Yas-Hiro Quano\raisebox{2mm}{\scriptsize 3 $\star$}
\end{large}
\vspace{24pt}
\begin{flushleft}
$~^1$
     \it Research Institute for Mathematical Sciences,
          Kyoto University, Kyoto 606-01, Japan  \\
      $~^2$
      \it Department of Mathematical Sciences,
          Faculty of Engineering Sciences, Osaka University,
          Toyonaka, \\
       Osaka 560, Japan \\
      $~^3$
      \it Department of Mathematics, University of Melbourne,
          Parkville, Victoria 3052, Australia
\end{flushleft}
\vspace{48pt}

\underline{ABSTRACT}
\end{center}
We consider the recently obtained integral representation of
quantum Knizhnik-Zamolodchikov equation of level 0.
We obtain the condition for the integral kernel
such that these solutions satisfy three axioms
for form factor \'{a} la Smirnov.
We discuss the relation between this
integral representation and the form factor of XXZ spin chain.
\vfill
\hrule

\vskip 3mm
\begin{small}

\noindent\raisebox{2mm}{$\star$} Supported by
the Australian Research Council.

\end{small}

\newpage

%\title{Form factors for the $XXZ$ model (Tentative)}

%\author{Takeo Kojima\raisebox{2mm}{\scriptsize 1},
%        Kei Miki\raisebox{2mm}{\scriptsize 2}
%    and Yas-Hiro Quano\raisebox{2mm}{\scriptsize 3}}

%\date{$~^1$
%     \it Research Institute for Mathematical Sciences,
%          Kyoto University, Kyoto 606-01, Japan \hfill \\
%      $~^2$
%      \it Department of Mathematical Sciences, Osaka University,
%          Toyonaka, Osaka 560, Japan \hfill \\
%      $~^3$
%      \it Department of Mathematics, University of Melbourne,
%          Parkville, Victoria 3052, Australia \hfill}
%\maketitle
%
%\begin{abstract}
%Form factors for $XXZ$ model is determined.
%\end{abstract}

\section{Introduction}

In \cite{JKMQ} an integral formula of the Smirnov type
was presented to the quantum Knizhnik--Zamolodchikov
($q$-KZ) equation \cite{FR}
of level $0$ associated with the vector representation
of the quantum affine algebra
$U_q \bigl( \widehat{\frak s \frak l _2}\bigr)$.
The $U_q \bigl( \widehat{\frak s \frak l _n}\bigr)$
generalization was studied in \cite{KQ}.
In these formulae, the freedom of solutions to $q$-KZ
equation corresponds to the choice of integral kernel
 with the cycle of integration being fixed.
The present paper is a step towards the determination of
the integral kernel given in \cite{JKMQ}
by studying the annihilation
pole structures of the solutions.

In the pioneering works \cite{Smbk},
Smirnov constructed the integral formulae of
form factors of the sine-Gordon model
that satisfy three axioms: (i) $S$ matrix symmetry;
(ii) (deformed) cyclicity; (iii) annihilation pole condition.
He utilized these axioms to construct
the matrix elements of local operators.
Refs \cite{JKMQ,KQ} were based on Smirnov's observation \cite{Sm1}
that (i) and (ii) imply the $q$-KZ equation of level $0$.
In these works,
instead of solving the $q$-KZ equation directly,
a system of difference equations
arising from
(i') the $R$ matrix symmetry and (ii) the deformed cyclicity
were considered.
At this moment the integral kernel of the formula
is arbitrary except that  it satisfies appropriate symmetries
and quasi-periodicity conditions.
These results can be easily modified so as to
enjoy  (i) instead of (i') for the $S$ matrix having the crossing
symmetry.
In this paper we shall derive the condition for the integral kernel
such that these solutions satisfy the third axiom for the
$U_q \bigl( \widehat{\frak s \frak l _2}\bigr)$ case.

The $q$-KZ equation was originally introduced
by Frenkel and Reshetikhin \cite{FR}
and was found to be  the master equation
for the  form factors
of solvable  lattice  models with the quantum affine symmetries
\cite{XXZ,JM}.
In this approach, the form factors can
be calculated by utilizing the  vertex operators of the algebra.
(See \cite{CORR,Kon} for  explicit calculations.)
Moreover, the form factors of
 appropriate
operators were shown to satisfy  Smirnov's
three axioms \cite{JM,Pak,M}.
Since our $S$ matrix coincides with the one
appearing in the XXZ model,
our solutions are expected  to be related
to the form factors of some
operators in the model.

The present paper is organized as follows.
In section 2 we formulate the problem and summarize our result.
In sections 3 and  4  we prove our result.
In section 5 we discuss our solutions in the context
of the $XXZ$ model.

\section{Problem and Result.}

The purpose of this section is
to formulate the problem, thereby
fixing our notations, and to state our result.

For a fixed complex parameter $q$ such that $0<|q|<1$,
let $U=U'_q \bigl(\widehat{\frak s \frak l _2}\bigr)$  be a $\bc$
algebra generated by $e_i$, $f_i$ and $t_i$ $(i=0,1)$ that satisfy
$$
[e_i,f_j]=\delta_{ij}{t_i-t_i^{-1}\over q-q^{-1}},\quad
t_i e_j t_i^{-1}=q^{4\delta_{ij}-2}e_j,
\quad t_i f_j t_i^{-1}=q^{2-4\delta_{ij}}f_j,\quad
t_it_j=t_j t_i,\quad t_i t_i^{-1}=t_{i}^{-1}t_i=1
$$
and the serre relations \cite{J}.
Let $\Delta$ be  the following coproduct of $U$
$$
\Delta(e_i)=e_i\otimes 1 +t_i\otimes e_i,\quad
\Delta(f_i)=f_i\otimes t_i^{-1} + 1\otimes f_i,\quad
\Delta(t_i)=t_i\otimes t_i,
$$
and set $\Delta' =\sigma \circ \Delta$, where
$\sigma (x\otimes y)=y\otimes x$ for $x,y \in U$.
Set $V\cong \bc v_+ \oplus \bc v_- $  and let $
(\pi_\zeta, V)\,$$(\zeta\in \bc\backslash \{0\})$
signify the vector representation of $U$ defined by
\begin{equation}
\begin{array}{l}
\pi_{\zeta} (e_1)(v_+ , v_-)=\zeta(0,v_+), \quad
\pi_{\zeta} (f_1) (v_+ , v_-)=\zeta^{-1}(v_-,0), \quad
\pi_{\zeta} (t_1) (v_+ , v_-)=(q v_+ , q^{-1} v_-), \\
\pi_{\zeta} (e_0)(v_+ , v_-)=\zeta(v_-,0), \quad
\pi_{\zeta} (f_0) (v_+ , v_-)=\zeta^{-1}(0,v_+), \quad
\pi_{\zeta} (t_0) (v_+ , v_-)=(q^{-1} v_+ , q v_-).
\end{array}
\label{eqn:pv}
\end{equation}
We shall later use the following abbreviation for its tensor
product representation via  $\Delta$
$$
 \pi_{(\zeta_1,\cdots,\zeta_N)}(y)
=(\pi_{\zeta_1}\otimes\cdots\otimes\pi_{\zeta_N})\circ
\Delta^{(N-1)}(y),\quad y\in U
$$

Let  $R(\zeta) \in \mbox{End}(V \otimes V)$
be the $R$ matrix of the six vertex model
$$
R(\zeta)v_{\varepsilon'_1}\otimes v_{\varepsilon'_2}=
\sum_{\varepsilon_1,\varepsilon_2}
v_{\varepsilon_1}\otimes v_{\varepsilon_2}
R(\zeta)^{\varepsilon_1 \varepsilon_2}
    _{\varepsilon'_1 \varepsilon'_2},
$$
where the nonzero entries are given by
$$
\begin{array}{l}
R(\zeta)^{++}_{++}=R(\zeta)^{--}_{--}=1,\\R
\displaystyle (\zeta)^{+-}_{+-}=R(\zeta)^{-+}_{-+}=b(\zeta)
={(1-\zeta^2)q\over 1-\zeta^2 q^2},\\
\displaystyle R(\zeta)^{+-}_{-+}=R(\zeta)^{-+}_{+-}=c(\zeta)
={(1-q^2)\zeta \over 1-\zeta^2 q^2 }.
\end{array}
$$
Then the following intertwining property holds \cite{J}:
\begin{equation}
R(\zeta_1 /\zeta_2)
(\pi_{\zeta_1} \otimes\pi_{\zeta_2})\circ\Delta(y)=
(\pi_{\zeta_1} \otimes\pi_{\zeta_2})\circ\Delta'(y)
R(\zeta_1 /\zeta_2).
\label{eqn:intertwine}
\end{equation}
We further introduce the scattering matrix  $S$ \cite{XXZ,JM}
$$
S(\zeta) =S_0 (\zeta) R(\zeta),
$$
where
$$
S_0(\zeta)=\zeta
\frac{(z^{-1};q^4)_{\infty}(zq^2 ;q^4)_{\infty}}
     {(z;q^4)_{\infty}(z^{-1}q^2 ;q^4)_{\infty}},
{}~~~~
(a;p_1,\cdots,p_n)_\infty=\prod_{k_i\ge 0}(1-a p_1^{k_1}\cdots p_n^{k_n}),
$$
and $z=\zeta^2$.
In what follows we shall work with
the tensor products of the vector space $V$'s.
Following the usual convention,
for $M \in \rm{End}(V)$ we let $M_j$
denote the operator on
$V^{\otimes N}$ acting as $M$
on the $j$-th tensor component and as identity on the other components.
Similarly for $X=S$ or $R$,
we let $X_{jk}(\zeta)$ ($j\neq k$)
signify the operator on
$V^{\otimes N}$ acting as $X(\zeta)$
on the $(j,k)$-th tensor
components and as identity
on the other components.
In particular we have
$X_{kj}(\zeta)=P_{jk}X_{jk}(\zeta)P_{jk}$, where
$P\in \mbox{End}(V\otimes V)$
stands for the transposition
$P(x\otimes y)=y\otimes x$.
We often use the following abbreviations
$$
\begin{array}{l}
X_{1,\cdots,N|N+1}(\zeta_1,\cdots,\zeta_N|\zeta_{N+1})=
X_{1,N+1}(\zeta_1/\zeta_{N+1})
\cdots X_{N,N+1}(\zeta_N/\zeta_{N+1}), \\
X_{N+1|1,\cdots,N}(\zeta_{N+1}|\zeta_1,\cdots,\zeta_N)=
X_{N+1,N}(\zeta_{N+1}/\zeta_N)
\cdots
X_{N+1,1}(\zeta_{N+1}/\zeta_1),
\end{array}
$$
where $X=S$ or $R$.

The main properties of $S(\zeta)$
are the Yang-Baxter equation
\begin{equation}
S_{12}(\zeta_1/\zeta_2)
S_{13}(\zeta_1/\zeta_3)
S_{23}(\zeta_2/\zeta_3)
=S_{23}(\zeta_2/\zeta_3)
S_{13}(\zeta_1/\zeta_3)
S_{12}(\zeta_1/\zeta_2),
\label{eqn:YBE}
\end{equation}
the initial condition
\begin{equation}
S(1)=-P,
\end{equation}
the unitarity relation
\begin{equation}
S_{12}(\zeta_1/\zeta_2)S_{21}(\zeta_2/\zeta_1)=1,
\label{eqn:unit}
\end{equation}
and the crossing symmetry
\begin{equation}
S_{12}(\zeta)v_\varepsilon\otimes u_\sigma=\sigma S_{31}(-\sigma/q\zeta)
v_\varepsilon\otimes u_\sigma
\label{eqn:cross}
\end{equation}
where $u_\sigma=v_{-}\otimes v_{+}+\sigma v_{+}\otimes v_{-}$
$~~(\sigma=\pm)$.

Let  $V^{(n l)}$ be the subspace of $V^{\otimes N}$ defined by
$$
V^{(n l)}
=\oplus~ \bc v_{\varepsilon_1} \otimes
\cdots \otimes v_{\varepsilon_N}
$$
where  the sum is taken over
$\varepsilon_j =\pm 1$ with fixed
$$
n=\sharp\{j\mid \varepsilon_j=-\}, \quad l
=\sharp\{j\mid \varepsilon_j=+\} \qquad(n+l=N)
$$
and let us consider a $V^{(nl)}$-valued function
$$
G^{(n l)}_{\varepsilon}(\zeta_1,\cdots,\zeta_N)=
\sum v_{\varepsilon_1} \otimes
\cdots \otimes v_{\varepsilon_N}
G^{(n l)}_{\varepsilon}(\zeta_1,\cdots,\zeta_N)
^{\varepsilon_1 \cdots \varepsilon_N}\qquad (\varepsilon=\pm).
$$

Our problem is  to obtain the function family
$G^{(n l)}_\varepsilon(\zeta_1,\cdots,\zeta_N)$
($\varepsilon=\pm,~ n,l=1,2,\cdots$) that
satisfy the following three axioms,

\noindent{\bf 1. $S$ matrix symmetry}
\begin{equation}
P_{j\,j+1} G_{\varepsilon}^{(n l)}
(\cdots,\zeta_{j+1},\zeta_j,\cdots)
\quad =
S_{j\,j+1}(\zeta_j/\zeta_{j+1})G_{\varepsilon}^{(n l)}
(\cdots,\zeta_j,\zeta_{j+1},\cdots)\qquad (1\leq j\leq N-1).
\label{eqn:S-symm} \\
\end{equation}

\noindent{\bf 2. Deformed cyclicity}
\begin{equation}
P_{12}\cdots P_{N-1 N} G_{\varepsilon}^{(n l)}
(\zeta_2,\cdots,\zeta_N, \zeta_1 q^{-2})
=
r_{\varepsilon}^{(l-n)}(\zeta_1) D^{(l-n)}_1 G_{\varepsilon}^{(n l)}
(\zeta_1,\cdots,\zeta_N).
\label{eqn:cyc}\\
\end{equation}
\noindent{\bf 3. Annihilation pole condition}
The $G_{\varepsilon}^{(n l)} (\zeta)$ has poles at
$\zeta_N =-\sigma \zeta_{N-1}/q$$\,\,(\sigma=\pm)$
 and the residue is given by
\begin{eqnarray}
&& \displaystyle
{\rm Res}_{\zeta_N/(-\sigma \zeta_{N-1}q^{-1})=1}
G_{\varepsilon}^{(n l)} (\zeta)\nonumber\\
&=&
\displaystyle\frac{1}{2}
\left( I-
\sigma^{N+1}r_{\varepsilon}^{(l-n)}(-\sigma \zeta_{N-1}q)
D^{(l-n)}_N S_{N-1|1,\cdots,N-2}
(\zeta_{N-1}|\zeta') \right)
G_{\sigma \varepsilon}^{(n-1 l-1)}(\zeta') \otimes u_{\sigma}.
\label{eqn:Res}
\end{eqnarray}
Here
$r_{\varepsilon}^{(k)}(\zeta)=\varepsilon \,r^{(k)}(\zeta)$,
 $r^{(k)}(\zeta)$ are scalar functions satisfying
\begin{equation}
r^{(k)}(\zeta)r^{(k)}(-\sigma \zeta q)=\sigma^N,
\label{eqn:rcond}
\end{equation}
$D^{(k)}=\mbox{diag }(\delta^{(k)}, \delta^{(k)}{}^{-1})$,
and $I$ is the identity operator.
In this paper we employ the convention
${\rm Res}_{x/y=1}G(x)=F(y)$
when $G(x)=\frac{1}{x/y -1}F(y)+O(1)$ at $x\simeq y$,
and we often use
the abbreviations
$(\zeta) =(\zeta_1 , \cdots , \zeta_N)$ and
$(\zeta') =(\zeta_1 , \cdots , \zeta_{N-2})$.
In section 5, we shall discuss the physical meaning of the above axioms
and the reason why we introduce $r^{(k)}(\zeta)$.

\noindent{\bf Remark}

\noindent
1. The action of the scattering matrix
$S(\zeta)$ preserves the values of $n$ and $l$.
The $S$ matrix further enjoys the $\bz _2$-symmetry. Therefore we can
consider $G_{\varepsilon}^{(n l)} (\zeta)\in V^{(nl)}$ and may
assume $n \leq l$ without loss of generality.

\noindent
2. The condition for $r^{(k)}(\zeta)$
(\ref{eqn:rcond}) follows from the consistency of the three axioms.
This can be seen from (\ref{eqn:rcond'}) and the argument below it.
Similarly we can show that
the diagonal operator $r^{(k)}_\varepsilon (\zeta)D^{(k)}$
depends only on $l-n$.

{}~

Hereafter we shall restrict ourselves to the case
$\delta^{(k)} =q^{-k/2}$ and $n\le l$,
since for this choice  the solutions to the
first two axioms were already obtained in \cite{JKMQ}.
In this paper, we shall derive the condition
for these solutions to satisfy the third axiom.
Before we state our result, we shall explain
how the solutions to the first two axioms are obtained from
the results of \cite{JKMQ}. Set
\begin{equation}
G_{\varepsilon}^{(nl)}(\zeta)
=\overline{G}_{\varepsilon}^{(nl)}(\zeta)
               \prod_{1\leq i< j \leq N} g(z_i/z_j)
\prod_{j=1}^N\zeta_j^{j-N},
\label{eqn:G}
\end{equation}
where
$$
g(z)={(z;q^4,q^4)_{\infty}(q^4/z;q^4,q^4)_{\infty}\over
(q^2z;q^4,q^4)_{\infty}(q^6/z;q^4,q^4)_{\infty}}.
$$
Then the first two axioms  can be recast as follows:
\begin{equation}
P_{j\,j+1} \overline{G}_{\varepsilon}^{(n l)}
(\cdots,\zeta_{j+1},\zeta_j,\cdots)
\quad =
R_{j\,j+1}(\zeta_j/\zeta_{j+1})
\overline{G}_{\varepsilon}^{(n l)}
(\cdots,\zeta_j,\zeta_{j+1},\cdots),
\label{eqn:R-symm}
\end{equation}
\begin{equation}
P_{12}\cdots P_{N-1 N}
\overline{G}_{\varepsilon}^{(n l)}
(\zeta_2,\cdots,\zeta_N, \zeta_1 q^{-2})
=
r_{\varepsilon} (\zeta_1)
D^{(l-n)}_1 \overline{G}_{\varepsilon}^{(nl)}
(\zeta_1,\cdots,\zeta_N) \prod_{j=2}^{N}
\frac{\zeta_j}{\zeta_1}.
\label{eqn:cyc'}
\end{equation}

Therefore, though the deformed cyclicity (\ref{eqn:cyc'})
is  different from
the one in \cite{JKMQ} by a multiplication factor,
$\overline{G}_{\varepsilon}^{(nl)}(\zeta)$ is similarly
shown to have the following
integral formula
\begin{equation}
\overline{G}_{\varepsilon}^{(nl)}(\zeta)
=
\displaystyle
\frac{1}{m!}\prod_{\mu =1}^{m} \oint_{C^{(N)}} \frac{dx_{\mu }}{2\pi i}
\Psi_{\varepsilon}^{(nl)}
(x_1 , \cdots , x_m | \zeta_1 , \cdots , \zeta_N )
\langle \Delta^{(nl)}  \rangle
(x_1,\cdots,x_m|\zeta_1,\cdots,\zeta_N ).
\label{eqn:G'}
\end{equation}
Here $m=n-1$ for $n=l$ and $m=n$ for $n<l$,
and $\Delta^{(nl)}(x_1,\cdots ,x_m | z_1,\cdots,z_n|z_{n+1},\cdots,z_N \bigr)$
 is the same polynomial obtained in \cite{JKMQ}.
(See (\ref{eqn:del1}) below, and
\cite{JKMQ} for further details.)
$\langle \Delta^{(nl)} \rangle
(x_1, \cdots , x_m| \zeta_1,\cdots,\zeta_N ) \in V^{(nl)}$
is the vector defined from this polynomial as follows
$$
\begin{array}{rcl}
\langle \Delta^{(nl)} \rangle
(x_1, \cdots , x_m| \zeta_1,\cdots,\zeta_N ) ^{-\cdots -+\cdots +}
&=&
\displaystyle\frac{\Delta^{(nl)} \bigl(
x_1, \cdots , x_m |
z_1, \cdots , z_n |z_{n+1}, \cdots , z_N \bigr)}
{\prod_{j=1}^n \prod_{i=n+1}^{N}(z_i -z_j \tau^2)}
\prod_{j=1}^{n} \zeta_j, \\
P_{j\,j+1} \langle \Delta^{(n l)} \rangle
(x_1, \cdots , x_m | \cdots,\zeta_{j+1},\zeta_j,\cdots)
&=&
R_{j\,j+1}(\zeta_j/\zeta_{j+1})
\langle \Delta^{(nl)} \rangle
(x_1, \cdots , x_m | \cdots,\zeta_j,\zeta_{j+1},\cdots),
\end{array}
$$
where and hereafter $\tau=q^{-1}$.
The integration $\oint_{C^{(N)}}dx_\mu$
is along a simple closed curve
$C^{(N)}=C^{(N)}(z_1, \cdots , z_N)$ oriented
anti-clockwise,
which encircles the points
$z_j \tau ^{-1-4k} (1\leq j \leq N , k\geq 0)$
but not
$z_j \tau ^{1+4k} (1\leq j \leq N , k\geq 0)$.
The kernel
$\Psi^{(nl)}_\varepsilon$
has the form
$$
\Psi^{(nl)}_{\varepsilon}
(x_1 , \cdots , x_m | \zeta_1 , \cdots , \zeta_N )=
\vartheta _{\varepsilon}^{(nl)}
(x_1 , \cdots , x_m | \zeta_1 , \cdots , \zeta_N )
\prod_{\mu =1}^{m}
\prod_{j=1}^{N} \psi \Bigl(\frac{x_{\mu }}{z_j }\Bigr).
$$
Here
$$
\psi(z)=\frac{1}{(zq;q^4)_{\infty}(z^{-1}q;q^4)_{\infty}}
$$
and $\vartheta^{(nl)}_\varepsilon$ is an arbitrary function that has
the following properties:
\begin{itemize}
\item
it is anti-symmetric and holomorphic
in the $x_\mu \in \bc \backslash \{0\}$,

\item
it is symmetric and meromorphic
in the $\zeta_j\in \bc \backslash \{0\}$,

\item
it has the two transformation properties
\begin{eqnarray}
&\displaystyle\frac
{\vartheta^{(nl)}_{\varepsilon} (x_1 , \cdots , x_m |
\zeta_1 , \cdots , \zeta_j \tau^2, \cdots, \zeta_N )}
{\vartheta^{(nl)}_{\varepsilon}
 (x_1 , \cdots , x_m | \zeta_1 , \cdots , \zeta_N )}
=
\tau^{N/2} r_{\varepsilon}^{(l-n)}(\zeta_j )
\displaystyle\prod_{\mu =1}^{m}
\frac{-z_j \tau}{x_{\mu }}
\prod_{k=1 \atop k \neq j}^{N} \frac{\zeta_k}{\zeta_j},
\label{eqn:zsym}\\
&\displaystyle \frac
{\vartheta^{(nl)}_{\varepsilon} (x_1 , \cdots, x_{\mu}\tau^4 ,
\cdots, x_m | \zeta_1 , \cdots , \zeta_N )}
{\vartheta^{(nl)}_{\varepsilon}
(x_1 , \cdots , x_m | \zeta_1 , \cdots , \zeta_N )}
=
\displaystyle\prod_{j=1}^{N} \frac{-x_{\mu } \tau}{z_j }.
\label{eqn:xsym}
\end{eqnarray}
\end{itemize}
Note that the first transformation function (\ref{eqn:zsym})
takes a form different from the one in \cite{JKMQ}  because of
the modification of (\ref{eqn:cyc'}).

Now we are in a position
to state the main result of this paper:

\begin{thm} The function family
$G_{\varepsilon}^{(nl)}(\zeta)$
defined in  (\ref{eqn:G}) and (\ref{eqn:G'}) satisfies
the annihilation pole condition (\ref{eqn:Res}) if
$\vartheta^{(nl)}_{\varepsilon}$
satisfies the following recurrence relation

\begin{equation}
\displaystyle \frac{\vartheta_{\varepsilon}^{(nl)}
(x_1 , \cdots ,x_{m-1}, z_{N-1}\tau |
\zeta_1 , \cdots , \zeta_{N-2},\zeta_{N-1}, -\sigma \zeta_{N-1} \tau^{-1})}
{\vartheta_{\sigma \varepsilon}^{(n-1,l-1)}
(x_1 , \cdots , x_{m-1} |
\zeta_1 , \cdots , \zeta_{N-2})}
=
\displaystyle c^{(nl)} z_{N-1}^{N-m-1}
\prod_{\mu =1}^{m-1}\theta(q x_{\mu}/z_{N-1}|q^2).
\label{eqn:rec}
\end{equation}

where
$$
\begin{array}{cl}
&\theta(z|q)=(z;q)_{\infty}(q/z;q)_{\infty},\\
&\displaystyle
c^{(nl)}=(-\tau)^{m-n-2l+3}{(q^2;q^2)_{\infty}^2}
\frac{(q^4;q^4,q^4)_{\infty}^2}{(q^2;q^4,q^4)_{\infty}^2}.
\end{array}
$$
\label{thm:main}
\end{thm}

There exists a nontrivial
example of the function $\vartheta_\varepsilon^{(nl)}$
which satisfies
the above conditions for a particular choice of $r^{(k)}(\zeta)$.
See section 5. We shall  check
the $(-\cdots -+\cdots +)$--component of
 the annihilation pole condition (\ref{eqn:Res}) in the next section
and show that Theorem \ref{thm:main} follows from it in section 4.

\section {The Annihilation Pole Condition for the Extreme Component}

In this section we show
the $(-\cdots -+\cdots +)$--component of
 the annihilation pole condition (\ref{eqn:Res}).
In terms of ${\overline G}_\varepsilon^{(nl)}$,
it can be recast as

\begin{prop}
If $\vartheta_\varepsilon^{(nl)}$ satisfies (\ref{eqn:rec}),
then  the following holds
\begin{equation}
\begin{array}{cl}
&\displaystyle
-2\frac{(q^2;q^4,q^4)_{\infty}^2}{(q^4;q^4,q^4)_{\infty}^2}
\frac{\sigma^{N+1}\tau^{(n-l)/2}
{\rm Res}_{\zeta_N /(-\sigma \zeta_{N-1}\tau)=1}
\overline{G}_{\varepsilon}^{(nl)} (\zeta)^{-\cdots -+\cdots +}}
{r_{\varepsilon}^{(l-n)}(-\sigma \zeta_{N-1}q) \zeta_{N-1}\prod_{j=1}^{N-2}
\left(\zeta_j\zeta_{N-1}\psi \left(\frac{\tau z_{N-1}}{z_j}\right) \right)
} \\
&=\displaystyle
\left(R_{N-1|1,\cdots, N-2}(\zeta_{N-1}|\zeta')
\overline{G}_{\sigma \varepsilon}^{(n-1 l-1)}
\otimes u_{\sigma}
\right)^{-\cdots -+\cdots +}.
\end{array}
\label{eqn:Res''}
\end{equation}
\label{prop:Res''}
\end{prop}

{\sl Proof}
{}From the definition of the $R$ matrix, the RHS is equal to
\begin{equation}
\sum_{k=1}^{n} c\left( \frac{\zeta_{N-1}}{\zeta_k} \right)
\prod_{j=k+1}^n b \left( \frac{\zeta_{N-1}}{\zeta_j} \right)
\overline{G}_{\varepsilon}^{(n-1,l-1)} (\zeta')^{
-\cdots \stackrel{k}{\check{+}}\cdots- \stackrel{n+1}{\check{+}}\cdots +}.
\label{eqn:a}
\end{equation}
Thanks to the $R$ symmetry (\ref{eqn:R-symm}),
we have the following relation
$$
\overline{G}_{\varepsilon}^{(n-1,l-1)} (\zeta')^{
-\cdots \stackrel{k}{\check{+}}\cdots -\stackrel{n+1}{\check{+}}\cdots +}\\
=
\displaystyle \prod_{j=k+1}^n
\frac{1}{b}\left( \frac{\zeta_k}{\zeta_j} \right)
\overline{G}_{\varepsilon,k}^{(n-1 l-1)}(\zeta')
-
\displaystyle \sum_{j=k+1}^n \frac{c}{b}
\left( \frac{\zeta_k}{\zeta_j} \right)
\prod_{i=k+1 \atop i \neq j}^n
\frac{1}{b}\left( \frac{\zeta_j}{\zeta_i} \right)
\overline{G}_{\varepsilon,j}^{(n-1 l-1)}(\zeta')
$$
where
$$
\overline{G}_{\varepsilon,j}^{(n-1,l-1)} (\zeta')
=
\overline{G}_{\varepsilon}^{(n-1 l-1)}
(\zeta_1, \stackrel{j}{\hat{\cdots}}, \zeta_n , \zeta_j, \zeta_{n+1}, \cdots ,
\zeta_{N-2})^{-\cdots-+\cdots+}\quad(1\le j\le n).
$$
Note that
(\ref{eqn:a}) contains only one term proportional to
$\overline{G}_{\varepsilon,1}^{(n-1 l-1)}(\zeta')$
when expressed in terms of $\overline{G}_{\varepsilon,j}^{(n-1 l-1)}(\zeta')$.
Since the result should be symmetric with respect to
$\zeta_1 , \cdots , \zeta_n$, we obtain
\begin{equation}
\mbox{RHS}
=
\displaystyle
\frac{(1-\tau^2)\zeta_{N-1}}{\prod_{j=1}^n(z_{N-1}-z_j\tau^2)}
\sum_{k=1}^n \zeta_k
\displaystyle \prod_{j=1 \atop j\neq k}^{n}
\frac{(z_{N-1}-z_j)(z_k-z_j \tau^2)}{z_k -z_j}
\displaystyle
\overline{G}_{\varepsilon ,k}
^{(n-1,l-1)} (\zeta').
\label{eqn:rhs}
\end{equation}

Let us turn to the  LHS.
In the calculation of  the residue of
$$
\overline{G}_{\varepsilon}^{(nl)}(\zeta)
^{-\cdots - +\cdots +}
=\frac{1}{m!}
\displaystyle \prod_{\mu =1}^{m}
\oint_{C^{(N)}} {dx_{\mu } \over 2\pi i}
\Psi_{\varepsilon}^{(nl)}
(x_1 , \cdots , x_m | \zeta )
\frac{\Delta^{(nl)}
(x_1, \cdots , x_m | z_1 , \cdots , z_n |
       z_{n+1}, \cdots , z_{N}) }
{\prod_{j=1}^{n}\prod_{i=n+1}^{N} (z_i -z_j \tau^2)}
 \prod_{j=1}^{n} \zeta_j ,
$$
at $\zeta_N =-\sigma \zeta_{N-1} \tau $,
we rewrite the integration as
\begin{equation}
\displaystyle\prod_{\mu =1}^{m}
\oint_{C^{(N)}} {dx_{\mu } \over 2\pi i}
=
\displaystyle \prod_{\mu =1}^{m-1}
\oint_{C'{}^{(N)}} {dx_{\mu } \over 2\pi i }
\left( \oint_{C'{}^{(N)}} {dx_m\over 2\pi i }
+m {\rm Res}_{x_m=z_N\tau^-1} \right),
\label{eqn:int+Res}
\end{equation}
in order to avoid the pinch of the contour $C^{(N)}$.
Here $C'{}^{(N)}$ is a simple anti-clockwise closed curve which
encloses the same poles but $z_N\tau^{-1}$ as for $C^{(N)}$,
and we used the symmetry of the integrand
with respect to $x_\mu$'s.
Since the integrand is regular at
$\zeta_N=-\sigma \zeta_{N-1} \tau$,
only the second term of the RHS of (\ref{eqn:int+Res})
contributes to  the residue.
Note that (\ref{eqn:rec}) is equivalent to
$$
\begin{array}{cl}
&\displaystyle
\frac{\vartheta_{\varepsilon}^{(nl)}
(x_1 , \cdots ,x_{m-1}, z_{N-1}\tau |
\zeta',\zeta_{N-1}, -\sigma \zeta_{N-1} \tau)}
{\vartheta_{\sigma \varepsilon}^{(n-1,l-1)}
(x_1 , \cdots , x_{m-1} |\zeta')}\\
&\displaystyle
= (-1)^{N-m+1}\tau^{3N/2-m-2}
c^{(nl)} \sigma^{N+1}r_\varepsilon^{(l-n)}(-\sigma \zeta_{N-1}q)
\prod_{j=1}^{N-2}\zeta_j\zeta_{N-1}
\prod_{\mu =1}^{m-1}\frac{\theta(q x_{\mu}/z_{N-1}|q^2)}{x_\mu},
\end{array}
$$
from (\ref{eqn:zsym}). We thus find
\begin{equation}
\begin{array}{cl}
\mbox{LHS}
&\displaystyle
=\frac {1}{(m-1)!}\prod_{\mu =1}^{m-1}
\oint_{C'^{(N)}} {dx_{\mu } \over 2\pi i}
\Psi_{\sigma \varepsilon}^{(n-1 l-1)}
(x_1,\cdots,x_{m-1}| \zeta')
\prod_{j=1}^n\prod_{i=n+1}^{N-1}\frac{1}{z_i-z_j\tau^2}\prod_{j=1}^n\zeta_j
\\
&
\displaystyle
\times (-1)^l\tau^{2-N}\zeta_{N-1}\frac
{\Delta^{(nl)}(x_1,\cdots,x_{m-1}, z_{N-1}\tau | z_1 , \cdots , z_n |
       z_{n+1}, \cdots ,z_{N-1}, z_{N-1}\tau^2)}
{\prod_{j=1}^n (z_{N-1}-z_j) \prod_{\mu=1}^{m-1} (x_{\mu}-z_{N-1}\tau)}.
\end{array}
\label{eqn:lhs}
\end{equation}

In order to show the equality of (\ref{eqn:rhs}) and (\ref{eqn:lhs}),
we shall prove
\begin{prop} The following holds:
\begin{equation}
\begin{array}{cl}
&
\displaystyle\frac
{\Delta^{(n l)}(x_1, \cdots , x_{m-1}, w\tau | z_1 , \cdots , z_n |
       z_{n+1}, \cdots , z_{N-2}, w, w\tau^2)}
{\prod_{j=1}^n (w-z_j)\prod_{\mu=1}^{m-1} (x_{\mu}-w\tau)}
=\tau^{n}
\sum_{k=1}^{n} \prod_{j=1 \atop j \neq k}^{n}
\frac{w-z_j}{z_k -z_j} \\
\times &
\displaystyle
\left\{ (-\tau)^{l-2}
(1-\tau^2) \displaystyle\prod_{i=n+1}^{N-2} (z_i -z_k \tau^2 )
\Delta^{(n-1\,l-1)}
(x_1, \cdots , x_{m-1}| z_1 , \stackrel{k}{\hat{\cdots}} ,
z_{n} |
z_k , z_{n+1}, \cdots , z_{N-2}) \right. \\
+ & \left.
\displaystyle\sum_{\nu =1}^{m-1} (-1)^{m+\nu}
h^{(N-2)}(x_{\nu}|z_1 , \cdots , z_{N-2})
\Delta'^{(m-2)}(x_{1}, \stackrel{\nu}{\hat{\cdots}}, x_{m-1}|
z_k|z_1 , \stackrel{k}{\hat{\cdots}},z_n|z_{n+1},\cdots, z_{N-2}) \right\},
\end{array}
\label{eqn:RecDel}
\end{equation}
Here
$h^{(N)}(x|z_1,\cdots,z_N)$ is the polynomial defined in \cite{JKMQ}
and
$$
\begin{array}{cl}
& \Delta'^{(m-2)}(x_{1}, \cdots, x_{m-2}|
z_1|z_2 ,\cdots,z_n|z_{n+1},\cdots, z_{N-2})\\
=&
{\rm det}\Bigl(
\left(A^{(n-1 l-1)}_{\lambda}(x_{\mu}|z'|z'')\right)
_{\scriptstyle{1\le \mu\le m-2} \atop
\scriptstyle{n-m+1 \le \lambda\le n-1}},
\left(f^{(n-1 l-1)}_{\lambda}(z_1 \tau | z'|z'')\right)_
{n-m+1\le \lambda\le n-1}\Bigr)
\end{array}
$$
where
$z'=(z_2,\cdots,z_n)$ and  $z''=(z_{n+1}, \cdots, z_{N-2},z_1)$.
\label{prop:restrict}
\end{prop}

{\sl Proof.}
Let us recall \cite{JKMQ} that
\begin{equation}
\Delta ^{(n l)}(x_1 , \cdots , x_m | z_1 , \cdots , z_n |
                      z_{n+1}, \cdots , z_N )
=\det \left( A_{\lambda }^{(n l)} (x_{\mu }|
z_1 , \cdots , z_n |
                      z_{n+1}, \cdots , z_N )
\right) _{\scriptstyle{1\le \mu\le m}
\atop\scriptstyle{n-m+1\le \lambda \le n}}
\label{eqn:del1}
\end{equation}
where
$$
\begin{array}{rl}
A_{\lambda }^{(n l)} (x| a_1 , \cdots , a_n |
b_1 , \cdots, b_l )= &
\displaystyle\prod_{j=1}^{n} (x-a_j \tau )
f_{\lambda }^{(n l)} (x| a_1 , \cdots , a_n | b_1 , \cdots ,b_l ) \\
& + \tau ^{2(l-n+\lambda -1)}
\displaystyle\prod_{i=1}^{l} (x-b_i \tau ^{-1})
g_{\lambda }^{(n)} (x| a_1 , \cdots , a_n ).
\end{array}
$$
As for the definition of
$f_{\lambda }^{(n l)}$ and $g_{\lambda }^{(n)}$,
see \cite{JKMQ}.
{}From
$$
A_{\lambda }^{(n l)} (w\tau | z_1 , \cdots , z_n |
z_{n+1} , \cdots, z_{N-2}, w, w\tau^2 )=
\displaystyle \tau^n \prod_{j=1}^{n} (w -z_j)
f_{\lambda }^{(n l)} (w\tau
| z_1 , \cdots , z_n | z_{n+1} , \cdots ,
z_{N-2}, w, w\tau^2 ),
$$
and the determinant structure of $\Delta^{(n l)}$,
we find that the LHS of (\ref{eqn:RecDel}) is  a polynomial in $w$.
{}From  the following properties ((4.3) and (4.15) in \cite{JKMQ})
$$
\begin{array}{l}
A_{\lambda }^{(n l)} (x| a_1 , \cdots , a_n |b_1 , \cdots, b_l )
\mbox{ is linear with respect to $b_i $'s,} \\
f_{\lambda }^{(n l)} (w \tau |
z_1,\cdots,z_n|z_{n+1},\cdots,z_{N-2}, w, w \tau ^2 )
\mbox{ is independent of $w$,}
\end{array}
$$
we further find that its degree is equal to $m-1$.
 Thus the LHS can be determined from $n(>m-1)$ values
at $w=z_k (1 \leq k \leq n)$.
For example, when $w=z_1$, using the recurrence relations of
$A^{(n l)}_\lambda$ and $f^{(n l)}_\lambda$ \cite{JKMQ},
its value is found to be
$$
\tau^n {\rm det}\Bigl(
\left(A^{(n-1 l-1)}_{\lambda}(x_{\mu}|z'|z'')\right)
_{\scriptstyle{1\le \mu\le m-1}
 \atop \scriptstyle{n-m+1 \le \lambda\le n}},
\left(f^{(n-1 l-1)}_{\lambda}(z_1 \tau | z'|z'')\right)_
{n-m+1\le \lambda\le n}\Bigr)
$$
where $z'$ and $z''$ are the same abbreviations as used
in the definition of $\Delta'{}^{(m-2)}$.
Then noting
$$
f^{(n-1 l-1)}_{n}(z_1 \tau |z'|z'')
=(-\tau)^{l-2}(1-\tau^2)\prod_{i=n+1}^{N-2}(z_i -z_1 \tau^2),
$$
we obtain (\ref{eqn:RecDel}).
{}~~~$\Box$

After the substitution of (\ref{eqn:RecDel}) into
(\ref{eqn:lhs})
we can replace
 $C'{}^{(N)}(z_1,\cdots,z_N)$ by $C^{(N-2)}(z_1,\cdots,z_{N-2})$.
Moreover, thanks to the same argument as
given in the proof of Lemma 3.2 in \cite{JKMQ},
the terms proportional
to $h^{(N-2)}(x_{\nu}|z_1, \cdots , z_{N-2})$
vanish after the integration.
Hence  Proposition \ref{prop:Res''} is proved.
{}~~~$\Box$

\section{Proof of the Theorem}

In this section we shall complete the proof of Theorem  \ref{thm:main}.
Firstly we show
\begin{prop}
{}~~~If the function family
$G^{(nl)}_{\varepsilon}(\zeta_1,\cdots,\zeta_N) \in V^{(nl)},
(\varepsilon=\pm,~n,l=1,2,\cdots)$ satisfies,

$1.~S-matrix~ symmetry~~ (\ref{eqn:S-symm}),$

$2.~Deformed~ cyclicity~~ (\ref{eqn:cyc}),$

$3.~ The~ (-\cdots-+\cdots+)-$component of
the $annihilation~ pole~ condition~
\it(\ref{eqn:Res}),$

$4.~\pi_{(\zeta_1,\cdots,\zeta_N)}(f_0)(G_{\varepsilon}^{(nl)}
(\zeta_1,\cdots,\zeta_N))=0,$

then the function family
$G^{(nl)}_{\varepsilon}(\zeta_1,\cdots,\zeta_N)$ satisfies
the $annihilation~ pole~ condition \it~(\ref{eqn:Res})$.
\label{prop:Y}
\end{prop}
{\sl Proof.}
In this proof, for $Y \in V^{\otimes N}$
we define $Y^{[\varepsilon_1 \varepsilon_2]}
\in V^{\otimes N-2}$ by
%\begin{equation}
%\displaystyle
$$
Y=\sum_{\varepsilon_1 \varepsilon_2}
Y^{[\varepsilon_1 \varepsilon_2]}\otimes v_{\varepsilon_1}
\otimes v_{\varepsilon_2},
$$
%\nonumber
%\end{equation}
and  call it the $[\varepsilon_1 \varepsilon_2]$ component of $Y$.

Let $A^{(nl)}_{\varepsilon,\sigma}(\zeta'|\zeta_{N-1})$ and
$B^{(nl)}_{\varepsilon,\sigma}(\zeta'|\zeta_{N-1})$
denote the LHS and RHS of the annihilation pole condition
(\ref{eqn:Res}), respectively, and set
$K^{(nl)}_{\varepsilon,\sigma}(\zeta'|\zeta_{N-1})
=A^{(nl)}_{\varepsilon,\sigma}(\zeta'|\zeta_{N-1})-
B^{(nl)}_{\varepsilon,\sigma}(\zeta'|\zeta_{N-1})$.
The aim of this proof is to show that $K^{(nl)}_{\varepsilon,\sigma}$
vanishes under the above four assumptions.
{}From the Yang--Baxter equation (\ref{eqn:YBE})
and the assumptions $1$ and $3$, we obtain
\begin{equation}
K^{(nl)}_{\varepsilon,\sigma}(\zeta'|\zeta_{N-1})^{[++]}=0.
\label{eqn:1}
\end{equation}

The intertwining property of the $S$ matrix implies
$$
S_{N-1|1,\cdots,N-2}(\zeta_{N-1}|\zeta')
(\pi_{\zeta'}\otimes \pi_{\zeta_{N-1}})
\circ \Delta'(y)
=(\pi_{\zeta'}\otimes \pi_{\zeta_{N-1}})
\circ \Delta(y)
S_{N-1|1,\cdots, N-2}(\zeta_{N-1}|\zeta'),\quad y\in U.
$$
Therefore  noting $G_\varepsilon^{(n-1\,l-1)}\in V^{(n-1\,l-l)}$
we find from the  assumption 4
\begin{equation}
\pi_{(\zeta', \zeta_{N-1}, -\sigma\zeta_{N-1}\tau)}(f_0)
K^{(nl)}_{\varepsilon,\sigma}(\zeta'|\zeta_{N-1})=0.
\label{eqn:2}
\end{equation}
Equations (\ref{eqn:1}) and (\ref{eqn:2}) imply that
$K^{(nl)}_{\varepsilon,\sigma}$
has the following form
$$
K^{(nl)}_{\varepsilon,\sigma}(\zeta'|\zeta_{N-1})
=K'^{(nl)}_{\varepsilon,\sigma}(\zeta'|\zeta_{N-1})\otimes u_\sigma.
$$

We shall show $K'{}^{(nl)}_{\varepsilon,\sigma}=0$ as follows.
The assumptions $1$ and $2$ give the following $q$-KZ equation
\begin{equation}
P_{N-1\,N}G_{\varepsilon}^{(nl)}(\zeta',\zeta_{N-1},\zeta_N)
=r_{\varepsilon}^{(l-n)}(\zeta_N q^2)D^{(l-n)}_{N-1}
S_{1,\cdots, N-2|N-1}(\zeta'|\zeta_N q^2)G_{\varepsilon}^{(nl)}
(\zeta',\zeta_N q^2,\zeta_{N-1}).
\label{eqn:rcond'}
\end{equation}

Taking the residue at $\zeta_N=-\sigma \zeta_{N-1}\tau$, we obtain
$$
P_{N-1\,N}A^{(nl)}_{\varepsilon,\sigma}(\zeta'|-\sigma\zeta_{N-1}\tau)
=- r_{\varepsilon}^{(l-n)}(\zeta_{N-1})D^{(l-n)}_{N-1}
S_{1,\cdots, N-2|N-1}(\zeta'|\zeta_{N-1})
A^{(nl)}_{\varepsilon,\sigma}(\zeta'|\zeta_{N-1}).
$$

{}From the unitarity and crossing symmetry of the $S$ matrix
(\ref{eqn:unit}),(\ref{eqn:cross}),
and the condition for $r^{(k)}(\zeta)$ (\ref{eqn:rcond}),
it follows that $B^{(nl)}_{\varepsilon,\sigma}(\zeta'|\zeta_{N-1})$ also
 satisfies the above equation. From these we find
$$
\sigma K'{}^{(nl)}_{\varepsilon,\sigma}(\zeta'|-\sigma\zeta_{N-1}\tau)
\otimes u_{\sigma}=-r_{\varepsilon}^{(l-n)}(\zeta_{N-1})D^{(l-n)}_{N-1}
S_{1,\cdots,N-2|N-1}(\zeta'|\zeta_{N-1})
K'{}^{(nl)}_{\varepsilon,\sigma}(\zeta'|\zeta_{N-1})\otimes
u_{\sigma}.
$$
By considering the difference of the $[-+]$ component and $\sigma\times$ the
$[+-]$ component
of the above equation, we obtain
$$
M^{(nl)}(\zeta'|\zeta_{N-1})
K'{}^{(nl)}_{\varepsilon,\sigma}(\zeta'|\zeta_{N-1})=0.
$$
Here
$$
M^{(nl)}(\zeta'|\zeta_{N-1})={\rm tr}_{N-1}\left( {\bar D}_{N-1}^{(l-n)}
R_{1,\cdots,N-2|N-1}(\zeta'|\zeta_{N-1})\right)
\in \mbox{End}(V^{\otimes N-2}),\quad
{\bar D}^{(k)}= \left(\matrix{1 &0\cr 0&-1}\right)D^{(k)}
$$
and  ${\rm tr}_{N-1}$ signifies the trace on the $(N-1)$-th space.
Since the matrix $M^{(nl)}$ is invertible for generic $\zeta_j$
(for example, consider the special case $\zeta_1=\cdots=\zeta_{N-1}$),
we obtain
$K'{}^{(nl)}_{\varepsilon,\sigma}(\zeta'|\zeta_{N-1})=0$.$~~~~\Box$

We have already shown that
the function family
$G^{(nl)}_{\varepsilon}(\zeta)$ defined in  section $2$
satisfies the assumptions 1--3 of Proposition \ref{prop:Y}.
Therefore  Theorem \ref{thm:main} is proved if we show
the following lemma.

\begin{lem}
\begin{equation}
(\pi_{(\zeta_1, \cdots, \zeta_N)}(f_0))
\langle \Delta^{(nl)} \rangle
(x_1, \cdots , x_m |\zeta_1,\cdots,\zeta_N ) =0.
\end{equation}
\label{lem:cop}
\end{lem}

{\sl Proof}
This lemma is proved similarly as Lemma 3 in Chap.7 in  \cite{Smbk}.
Set
$$
\begin{array}{cl}
&P^{(n-1\,l+1)}(x_1,\cdots,x_m|z_1,\cdots,z_{n-1}|z_n,\cdots,z_N) \\
=&(-\tau)^l\displaystyle\prod_{j=1}^{n-1}\zeta_j^{-1}
\prod_{i=n}^N\prod_{j=1}^{n-1}(z_i-z_j\tau^2)
\bigl(\pi_{(\zeta_1,\cdots,\zeta_N)}
(f_0)\langle \Delta^{(n l)}\rangle (x_1,\cdots,x_m|\zeta_1,\cdots,\zeta_N)
\bigr)^{-\cdots -+\cdots +}.
\end{array}
$$
Then thanks to the $R$ symmetry we obtain
$$
\begin{array}{cl}
&P^{(n-1\,l+1)}(x_1,\cdots,x_m|z_1,\cdots,z_{n-1}|z_n,\cdots,z_N)\\
=&\displaystyle\sum_{k=n}^N
\frac{\prod_{j=1}^{n-1}(z_k-z_j\tau^2)}
{\prod_{\scriptstyle{i=n}\atop \scriptstyle{i\ne k}}^N (z_k-z_i)}
\Delta^{(n l)}(x_1,\cdots,x_m|z_1,\cdots,z_{n-1},z_k|z_n,
\mathop{\hat{\cdots}}^{k},z_N).
\end{array}
$$
{}From  the recurrence relation of $\Delta^{(nl)}$ \cite{JKMQ},
we can see that $P^{(n-1\,l+1)}$ satisfies
$$
\begin{array}{l}
P^{(n-1\,l+1)}(x_1,\cdots,x_m|z',a|z'',a\tau^2)=
\displaystyle\prod_{\mu=1}^m(x_\mu-a\tau)
\sum_{\nu=1}^m(-1)^{m+\nu}h^{(N-2)}(x_\nu|z',z'')
P^{(n-2\,\,l)}(x_1,\mathop{\hat{\cdots}}^\nu,x_m|z'|z'') \\
\end{array}
$$
where $z'=(z_1,\cdots,z_{n-2})$ and $z''=(z_{n},\cdots,z_{N-1})$.
We further find from the properties of  $\Delta^{(nl)}$
as a polynomial \cite{JKMQ}
that  $P^{(n-1\,l+1)}$ is a
homogeneous polynomial of degree $\displaystyle{{m \choose 2}}+(n-1)l-1$,
antisymmetric with respect to $x_\mu$'s and symmetric
with respect to
$\{z_1,\cdots,z_{n-1}\}$ and $\{z_{n},\cdots,z_N\}$, respectively.
Moreover from the power counting  $P^{(0\,l+1)}=0$.
{}From these properties we can show $P^{(n-1l+1)}=0$.
 ~~~~$\Box$

\section{Discussion}

In this section we discuss the function family $G^{(nl)}_\varepsilon(\zeta)$
satisfying the three axioms (\ref{eqn:S-symm}--\ref{eqn:Res})
in the context of the $XXZ$ model in the ferroelectric regime ($-1<q<0$).
In refs \cite{XXZ,JM}, this model was solved
by utilizing the representation theory of  $U_q (\widehat{\frak s \frak l_2})$.
Firstly following mainly their notations,
we  summarize the necessary results on this model.
See those references for further details.
Let  $V(\Lambda_i)$ $(i=0,1)$ be the level $1$ highest weight module
of $U_q (\widehat{\frak s \frak l_2})$ and set
${\cal  H}=V(\Lambda_0)\oplus V(\Lambda_1)$ and
${\cal F}^{(ij)}=V(\Lambda_i)\otimes V(\Lambda_j)^{*b}$.
Here the superscript $*b$ signifies  the dual module
regarded as a left module by some anti-automorphism $b$.
Then the  space ${\cal F}$ on which the $XXZ$ hamiltonian acts
is identified with
$$
{\cal F}={\cal H}\otimes {\cal H}^{*b}=
\bigoplus_{i,j=0,1}{\cal F}^{(ij)}.
$$
In this space ${\cal F}$
there exist  two ground states of the hamiltonian,  which
belong to  ${\cal F}^{(00)}$ and ${\cal F}^{(11)}$. We denote them and
their dual vectors by $|{\rm vac}\rangle _{(i)}$ and
${}_{(i)}\langle{\rm vac}|$  $(i=0,1)$, respectively.
The creation and annihilation operators
$
\varphi^*_{\varepsilon}(\zeta),
\varphi^{\varepsilon}(\zeta) \in
\oplus_{ij}{\rm Hom}({\cal F}^{(ij)},{\cal F}^{(1-i\,j)})
$
$(\varepsilon=\pm,|\zeta|=1)$
that diagonalize the hamiltonian can be constructed
in terms of  the vertex operators of the algebra.
They have the following property
$$
\varphi^\varepsilon(\zeta)|{\rm vac}\rangle _{(i)}=0,\qquad
{}_{(i)}\langle{\rm vac}|\varphi^*_\varepsilon(\zeta)=0,
$$
and the whole space ${\cal F}$ and its dual space
${\cal F}^*$ are spanned by the vectors
$$
\varphi^*_{\varepsilon_1}(\zeta_1)\cdots
\varphi^*_{\varepsilon_N}(\zeta_N)|{\rm vac}\rangle _{(i)}
\quad\mbox{and}\quad
{}_{(i)}\langle{\rm vac}|
\varphi{}^{\varepsilon_1}(\zeta_1)\cdots
\varphi{}^{\varepsilon_N}(\zeta_N)
\quad (i=0,1,\, N=0,1,\cdots)
$$
respectively.
These operators further  satisfy the following property
\begin{equation}
\varphi^*_\varepsilon(-\zeta)=(-1)^{(\varepsilon-(-1)^i)/2}
\varphi^*_\varepsilon(\zeta),\quad
\varphi^\varepsilon(-\zeta)=(-1)^{(\varepsilon+(-1)^i)/2}
\varphi^\varepsilon(\zeta)\quad \mbox{on   }{\cal F}^{(ij)}\quad (i,j=0,1)
\label{eqn:pr}
\end{equation}
and the commutation relations
$$
\begin{array}{cl}
&P_{12}\varphi^V(\zeta_1)\varphi^V(\zeta_2)=S_{12}(\zeta_1/\zeta_2)
\varphi^V(\zeta_2)\varphi^V(\zeta_1),\quad
P_{12}\varphi^*{}^V(\zeta_1)\varphi^*{}^V(\zeta_2)=S_{12}(\zeta_1/\zeta_2)
\varphi^*{}^V(\zeta_2)\varphi^*{}^V(\zeta_1),\\
&P_{12}\varphi^V(\zeta_1)\varphi^*{}^V(\zeta_2)=
S_{12}(-\zeta_1/q\zeta_2)
\varphi^*{}^V(\zeta_2)\varphi^V(\zeta_1)+
\displaystyle\frac{1}{2}
\sum_{\sigma=\pm} \sigma\delta(\sigma\zeta_1/\zeta_2)
u_\sigma\otimes (P_0+\sigma P_1).
\end{array}
$$
In the above equations,
$$
\varphi^*{}^V(\zeta)=\sum_\varepsilon v_\varepsilon \otimes
\varphi^*_{-\varepsilon}(\zeta), \qquad
\varphi^V(\zeta)= \sum_\varepsilon  v_\varepsilon \otimes
\varphi^{\varepsilon}(\zeta) \in
\bigoplus_{ij}{\rm Hom}({\cal F}^{(ij)},V\otimes{\cal F}^{(1-i\,j)}),
$$
$P_i$ is the projection operator to the subspace
$\oplus_{j=0,1} {\cal F}^{(ij)}$
and $\delta(\zeta)=\sum_{m\in\bz}\zeta^m$.
Note that we consider the creation and annihilation operators
in the principal picture. (See, for example,
eqs.(2.2, 4.8-9) in \cite{M} for the above properties of these operators
in this picture.)

Now we shall discuss the form factors of this model.
For simplicity, we shall consider an operator
${\cal O}\in {\rm End}({\cal F})$ that
has the form $id\otimes O $ ( $O\in {\rm End}({\cal H}^*)$) and satisfies
\begin{equation}
(id \otimes {\bar \Psi}^V(\zeta)){\cal O}=r(-\zeta q)D
{\cal O}(id \otimes {\bar \Psi}^V(\zeta)).
\label{eqn:con}
\end{equation}
Here $r(\zeta)$ is a scalar function,
$D$ is a diagonal matrix acting on the space $V$
and ${\bar \Psi}^V(\zeta):{\cal H}^{*b}\to{\cal H}^{*b}\otimes V_\zeta$
is the type II vertex operator in the terminology of \cite{XXZ,JM}.
We denoted by $V_\zeta$ the vector representation
defined in (\ref{eqn:pv}).
For such ${\cal O}$ we introduce the form factor by
$$
G^{(N)}_\varepsilon(\zeta_1,\cdots,\zeta_N)
=\sum_{i=0,1} \varepsilon^i \times {}_{(i)}\langle{\rm vac}|
{\cal O}\varphi^*{}^V(\zeta_N)\cdots \varphi^*{}^V(\zeta_1)
|{\rm vac}\rangle_{(i+N)}\quad (\varepsilon=\pm1).
$$
The commutation relations among the creation operators
imply the first axiom. From (\ref{eqn:pr}), we have
\begin{equation}
G^{(N)}_{\varepsilon}(\zeta_1,\cdots,-\zeta_j,\cdots,\zeta_N)
^{-\cdots-+\cdots+}
=(-1)^{N-j+H(n\ge j)}G^{(N)}_{-\varepsilon}(\zeta_1,\cdots,
\zeta_j,\cdots,\zeta_N)^{-\cdots-+\cdots+},
\label{eqn:Gpr}
\end{equation}
where $H$ is the step function and $n$ is the number of the superscript $-$
of $G^{(N)}_\varepsilon$.
Note that for  $G^{(nl)}_\varepsilon(\zeta)$ defined in section 2
the condition (\ref{eqn:Gpr}) is equivalent to
\begin{equation}
\vartheta^{(nl)}_{\varepsilon}(x_1,\cdots,x_m|\zeta_1,\cdots,
-\zeta_j,\cdots,\zeta_N)=
\vartheta^{(nl)}_{-\varepsilon}
(x_1,\cdots,x_m|\zeta_1,\cdots,\zeta_j,\cdots,\zeta_N).
\label{eqn:thetacon}
\end{equation}
Let   $\zeta^{\pm}$ signify $(1\pm\eta)\zeta$ $\,\,(0<\eta\ll 1)$
for $\zeta$,
and  $\zeta''$ be the abbreviation $(\zeta_1,\cdots,\zeta_{N-1})$.
Set  $r_\varepsilon(\zeta)=\varepsilon r(\zeta)$ as before and
set further
$$
G^{(N-2)}_{\varepsilon,\sigma,k}(\zeta_1,\cdots,\zeta_{N-1})=
P_{k,k+1}\cdots P_{N-2,N-1}
G^{(N-2)}_{\sigma\varepsilon}(\zeta_1,\stackrel{k}{\hat{\cdots}},\zeta_{N-1})
\otimes u_\sigma.
$$
Then similarly  as in \cite{M},
thanks to  the condition (\ref{eqn:con})
the following matrix element of ${\cal O}$
can be expressed
in terms of the form factors in two different ways:
\begin{equation}
\begin{array}{cl}
&\displaystyle\sum_{i=0,1} \varepsilon^i \times
{}_{(i)}\langle{\rm vac}|\varphi^V(\zeta_N){\cal O}
\varphi^*{}^V(\zeta_{N-1})\cdots \varphi^*{}^V(\zeta_1)
|{\rm vac}\rangle_{(i+N)}\\
=&\displaystyle G^{(N)}_\varepsilon(\zeta'',\zeta_N^{-}/(-q))
+\frac{1}{2}\sum_{\scriptstyle 1\le k \le N-1\atop
\scriptstyle \sigma=\pm}\delta(\sigma\zeta_k/\zeta_N)
S_{k+1,\cdots,N-1|k}(\zeta_{k+1},\cdots,\zeta_{N-1}|\zeta_k)
G^{(N-2)}_{\varepsilon,\sigma,k}(\zeta'')\\
=& r_\varepsilon(-\zeta_N q)D_N \Bigl(
P_{N-1,N}\cdots P_{1,2}
G^{(N)}_\varepsilon(-q\zeta_N^{+},\zeta'')\\
&\hskip3mm+\displaystyle\frac{1}{2}
\sum_{\scriptstyle 1\le k \le N-1\atop
\scriptstyle \sigma=\pm}\sigma^{N+1}
\delta(\sigma\zeta_k/\zeta_N)
S_{k|1,\cdots,k-1}(\zeta_k|\zeta_1,\cdots,\zeta_{k-1})
G^{(N-2)}_{\varepsilon,\sigma,k}(\zeta'') \Bigr).
\label{eqn:mat}
\end{array}
\end{equation}
The last equality of (\ref{eqn:mat})
implies that $G^{(N)}_\varepsilon(\zeta)$ satisfies
also the second and third axioms.
Conversely,
suppose that  we are given the $G^{(nl)}_\varepsilon(\zeta)$
satisfying the three axioms and (\ref{eqn:Gpr}).
Then we can define an operator ${\cal O}\in {\rm End}({\cal F})$
by giving the matrix elements in terms of
$G^{(nl)}_\varepsilon(\zeta)$ in two different ways
as in the original work \cite{Smbk}.
The second and third axioms assures the equivalence of
the two expressions.
(Equation (\ref{eqn:mat}) shows one simplest example.)
These two expressions
enable us to calculate the commutation relations of the thus defined
operators, though the knowledge of the poles other than
annihilation poles of the $G^{(nl)}_\varepsilon(\zeta)$  is necessary.
The classification of all solutions to the conditions
(\ref{eqn:zsym}--\ref{eqn:rec}, \ref{eqn:thetacon}) and
the identification of them with ${\cal O} \in {\rm End}({\cal F})$
are still open questions.

Finally we shall give an example of  $\vartheta^{(nl)}_\varepsilon$
in the case
$$
r^{(nl)}_\varepsilon(\zeta)=\varepsilon \prod_{k=1}^{N-2m} t(\zeta/\xi_k),
\quad
t(\zeta)=\zeta^{-1}{\theta(q z|q^4)\over\theta(q z^{-1}|q^4)},
$$
where $m=n-1$ for $n=l$$,=n$ for $n<l$ as before and $\xi_k\in {\bc}$.
In this case one solution to the conditions (\ref{eqn:zsym}--\ref{eqn:rec}
, \ref{eqn:thetacon}) is given by
\begin{equation}
{\vartheta^{(n l)}_\varepsilon (x_1,\cdots,x_m|\zeta_1,\cdots,\zeta_N)\over
\prod_{j=1}^N\zeta_j^{N-2m}
\prod_{1\leq \mu < \nu \leq m}\left(x_\mu\theta(x_\nu/x_\mu|q^2)\right)}
={\rm const}\,\varepsilon^N
\theta(-\varepsilon u|q^2)
\prod_{\scriptstyle 1\le k \le N-2m \atop
\scriptstyle 1 \le \mu \le m}\theta(\xi_k^2/x_\mu|q^4)
\prod_{\scriptstyle 1\le j\le N \atop
\scriptstyle 1\le k \le N-2m}f(z_j/\xi_k^2),
\end{equation}
where
$$
u=(-1)^{N-m}\tau^{N/2-3m}
{\prod_{\mu=1}^m x_\mu \prod_{k=1}^{N-2m}
\xi_k \over \prod_{j=1}^N\zeta_j},\quad
f(z)={(q^5z;q^4,q^4)_\infty(q^5/z;q^4,q^4)_\infty \over
(q^3z;q^4,q^4)_\infty(q^3/z;q^4,q^4)_\infty}.
$$
{}~

\section*{Acknowledgments}
The authors would like to thank M. Jimbo and T. Miwa
for the earlier collaborations. Y.-H.Q. is supported by
the Australian Research Council.

\end{document}